\documentclass[a4paper,10pt]{article}
\usepackage[dvips]{graphicx}
\usepackage{amssymb,amsmath}
\numberwithin{equation}{section}

\begin{document}

\title{Dark energy in  some integrable and nonintegrable FRW cosmological models}
\vspace{4cm}
\author{Kuralay  Esmakhanova$^1$,  Nurgissa Myrzakulov$^1$, Gulgasyl Nugmanova$^1$, \\
 Yerlan Myrzakulov$^1$, Leonid Chechin$^1$,   Ratbay Myrzakulov$^{1,2}$\footnote{The corresponding author. Email: rmyrzakulov@gmail.com; rmyrzakulov@csufresno.edu}\vspace{1cm}\\ \textit{$^1$Eurasian International Center for Theoretical Physics,} \\ \textit{Eurasian National University, Astana 010008, Kazakhstan}\\ \textit{$^2$Department of Physics, CSU Fresno, Fresno, CA 93740 USA}}

\date{}

\maketitle
\vspace{2cm}
\begin{abstract} 

One of the greatest challenges in cosmology today is to determine  the nature of dark energy, the sourse of the observed present acceleration of the Universe.  Besides the vacuum energy, various dark energy models have been suggested.  The Friedmann - Robertson - Walker (FRW) spacetime plays an important role in modern cosmology. In particular, the most popular models of dark energy work in the FRW spacetime.   In this work, a new class of integrable  FRW cosmological models is presented. These models induced by the well-known Painlev$\acute{e}$ equations. Some nonintegrable FRW models are also considered. These last models are constructed with the help of Pinney, Schr$\ddot{o}$dinger and hypergeometric equations. Scalar field description and two-dimensional generalizations of some  cosmological  models are presented. Finally some integrable and nonintegrable $F(R)$ and $F(G)$ gravity models are constructed. 
\end{abstract}
\sloppy
\vspace{2cm}
\tableofcontents
\vspace{2cm}
\section{Introduction} 
Recent measurements of redshift and luminosity-distance relations of type $Ia$ supernovae indicate that the expansion of the universe is accelerating \cite{Perlmutter}-\cite{Riess}. This appears to be in strong disagreement with the standard picture of a matter dominated universe. These observations can be accommodated theoretically by postulating that certain exotic matter (dark energy)  with negative pressure domainates the present epoch of our universe. Understanding the nature of dark energy and its many related problems is  one of the most formidable challenges  in modern cosmology (see e.g. \cite{Nojiri1}-\cite{Dzhun}).  Almost all models of dark energy work in the FRW spacetime that is in other words are the FRW models.  Practically, all FRW cosmological models of dark energy face some difficults related with some nontrivial problems like the coincidence problem, the fine-turn problem and so on. Such and other problems of modern cosmology demand more carefully investigate the physical and mathematical nature of the FRW models.  In this context, it is important to study integrable (and nonintegrable) cases of the FRW models.  Here we mention  the  well-known fact that Einstein's gravitational equations and their some  generalizations admit  integrable reductions (see e.g. Refs. \cite{Maison}-\cite{Alekseev2} and references therein).  In this work, we consider some examples of  integrable and nonintegrable FRW cosmological models induced by some known linear and nonlinear second-order  ordinary differential equations (ODE) (see also \cite{Kuralay}-\cite{Nugmanova2}). 

The paper is organized as follows. In section 2, we give some main informations on the FRW cosmological model for the flat spacetime case. In the next section, we study
some nonintegrable FRW cosmological models. A new class of integrable FRW models was considered in section 4. These results were generalized for the modified $f(R)$ and $F(G)$ gravity models in sections 5 and 6. Sections 7 and 8 are devoted to the scalar field description and two-dimensional generalizations of some models. In the last section, we give the conclution.

\section{ FRW cosmology} 
We start with the standard gravitational action
  \begin{equation}
S=\int\sqrt{-g}d^4x (R+L_m). \end{equation}
Here $R$ is the scalar curvature and $L_m$ is the Lagrangian of the matter. 
Now we consider the FRW spacetime with the scale factor $a(t)$ and metric
   \begin{equation}
ds^2=-dt^2+a^2(t)\left[\frac{dr^2}{1-kr^2}+r^2d\Omega^2\right]. \end{equation}
Here $k$ can take any value but it is related to $(-, 0, +)$ curvatures according to sign.  One sets $\dot{a}=da/dt$ and computes Christoffel symbols to be
$$
\Gamma^0_{11}=\frac{a\dot{a}}{1-kr^2},\quad \Gamma^1_{11}=\frac{kr}{1-kr^2}, \quad \Gamma^0_{22}=a\dot{a}r^2,\quad \Gamma^0_{33}=a\dot{a}r^2\sin^2{\theta},
$$
 \begin{equation}
\Gamma^1_{01}=\Gamma^2_{02},\quad \Gamma^3_{03}=\frac{\dot{a}}{a}, \quad \Gamma^1_{22}=-r(1-kr^2),\quad \Gamma^1_{33}=-r(1-kr^2)\sin^2{\theta}, \end{equation}
$$
\Gamma^2_{12}=\Gamma^3_{13}=r^{-1},\quad \Gamma^2_{33}=-\sin{\theta}\cos{\theta}, \quad \Gamma^3_{23}=\cot{\theta},
$$
 They lead to
 $$
 R_{00}=-3\frac{\ddot{a}}{a},\quad R_{11}=\frac{a\ddot{a}+2\dot{a}^2+2k}{1-kr^2},
 $$
  \begin{equation}
  R_{22}=r^2[a\ddot{a}+2\dot{a}^2+2k], \quad R_{33}=r^2[a\ddot{a}+2\dot{a}^2+2k]\sin^2{\theta}
  \end{equation}
  with Ricci scalar 
   \begin{equation}
  R=6\left[\frac{\ddot{a}}{a}+\left(\frac{\dot{a}}{a}\right)^2+\frac{k}{a^2}\right].  \end{equation}
 The Friedmann equations read as
  \begin{equation}
  \left(\frac{\dot{a}}{a}\right)^2=\frac{8\pi G}{3}\rho-\frac{k}{a^2}, \quad
  \frac{\ddot{a}}{a}=-\frac{4\pi G}{3}(\rho+3p),  \end{equation}
 where we note that $p<-\rho/3$ implies repulsive gravitation. Recaling the Hubble parameter $H=\dot{a}a^{-1}$  these equations can be written as
 \begin{equation}
  H^2=\frac{8\pi G}{3}\rho-\frac{k}{a^2}, \quad
  \dot{H}=-4\pi G(\rho+3p)+\frac{k}{a^2}.
    \end{equation} The  Friedman equations with a cosmological constant for a homogeneous isotropic universe models  are
  \begin{equation}
  \left(\frac{\dot{a}}{a}\right)^2=\frac{\Lambda}{3}-\frac{k}{a^2}+\frac{8\pi G}{3}\rho, \quad
  \frac{\ddot{a}}{a}=\frac{\Lambda}{3}-\frac{4\pi G}{3}(\rho+3p),  \end{equation}
 In this work, we consider the case: $k=\Lambda=0$ and set $8\pi G=1$. 
If the FRW  spacetime is filled with a fluid of energy density $\rho$ and pressure $p$, then conservation of energy-momentum tensor 
 \begin{equation}
\nabla^{\mu}T_{\mu\nu}=0, \end{equation}
gives
 \begin{equation}
\dot{\rho}+3H(\rho+p)=0, \end{equation}
where a dot represents differentiation with respect to $t$.
 So finally the equations for the action (2.1) we can write in   \textit{the H-form}
	\begin{eqnarray}
		p&=&-2\dot{H}-3H^2,\\
	\rho&=&3H^2,\\
	\dot{\rho}&=&-3H(\rho+p)
	\end{eqnarray}
	or in  \textit{the N-form}	\begin{eqnarray}
		p&=&-2\ddot{N}-3\dot{N}^2,\\
	\rho&=&3\dot{N}^2,\\
	\dot{\rho}&=&-3\dot{N}(\rho+p),
	\end{eqnarray}
	where $N=\ln{a}$. In this work, we study some FRW cosmogical models, for example,  with the equation of state (EoS) \cite{Nojiri1}-\cite{Nojiri3} (see e.g.  also \cite{Elizalde}-\cite{Dzhun})
\begin{equation}
		p=f_1(\dot{N}, N,t)\rho+f_{2}(\dot{N}, N,t)\rho^{\beta}+f_3(\dot{N}, N,t),
	\end{equation}
	where $ \beta\neq 1$ and  $f_i=f_i(\dot{N}, N,t)$ are some
	 functions of $\dot{N}, N$ and $t$.
\section{Nonintegrable FRW cosmological models} 
In this section, we consider some known and new FRW models, for which, $N$ satisfies some second-order linear and nonlinear ODEs. These ODEs are   nonintegrable so that the corresponding FRW 
cosmological models are nonintegrable (see e.g. \cite{Kuralay}-\cite{Nugmanova2}). Before we go any further, we explain more clearly (as mathematical claim) what is integrable FRW cosmology. If one of dependent variables of some  FRW model, say $N$ for our case,  satisfies some integrable differential equation then such FRW model we call an \textit{integrable FRW model} (or in other words \textit{integrable FRW cosmology}). On the other hand, if $N$ satisfies some nonintegrable differential  equation then such FRW model we call a \textit{nonintegrable FRW model} (or in other words \textit{nonintegrable FRW cosmology}). Now let us give the definition of integrable equations. In mathematics and physics, there are various distinct notions that are referred to under the name of integrable systems (equations). In our case, the  given nonlinear differential equation is said to be \textit{integrable} if it admits infinite number of independent conserved quantities (that is integrals of motion)  ($I_1, I_2, ...$) in involution. Two conserved quantities ($I_i, I_j$) are said to be in involution if they Poisson-commute that is $\{I_i, I_j\}=0$ where $\{f,g\}$ is some Poisson bracket. 
 
\subsection{$\Lambda$CDM cosmology}
We start with the $\Lambda$CDM cosmology. Let the parametric EoS has the form
\begin{equation}
		p=-\Lambda,\quad 	\rho=3\dot{N}^2.
	\end{equation}
It is well-known that in this case,  $N$ satisfies the following  
equation 
\begin{equation}
 \ddot{N}=0.5\Lambda-1.5\dot{N}^2.\end{equation}
 The corresponding EoS parameter reads as
 \begin{equation}
 \omega=-1+\frac{\rho_0}{\rho_0+\Lambda a^3}.\end{equation}

 \subsection{Pinney  cosmology}
 In this subsection, we study the cosmological models 
  induced by the Pinney equation. This equation we here write  as  
 \cite{Ermakov} - \cite{Pinney} (see also e.g. \cite{Kamenshchik})
  \begin{equation}
 \ddot{y}=\xi(t)y+\frac{k}{y^3},\end{equation}
 where $\xi=\xi(t),\quad k=const$.

1) Let the parametric EoS has the form
\begin{eqnarray}
		p&=&-3\dot{N}^2+2\zeta^2(t)N-2kN^{-3},\\
	\rho&=&3\dot{N}^2.
	\end{eqnarray}
Then it can be shown that $N$ satisfies the Pinney equation
\begin{equation}
 \ddot{N}=-\zeta^2(t)N+\frac{k}{N^3},\end{equation}
 where $\zeta=\zeta(t),\quad k=const$. It is known that as $\zeta=1$, 
 the Pinney  equation has the following  particular solution (see e.g. \cite{Haas})
 \begin{equation}
 N=(\cos^2{t}+k\sin^2{t})^{0.5}.\end{equation}
Then
\begin{eqnarray}
 \dot{N}&=&(k-1)\sin{t}\cos{t}(\cos^2{t}+k\sin^2{t})^{-0.5}, \\ \ddot{N}&=&[k-(\cos^4{t}+2k\sin^2{t}\cos^2{t}+k^2\sin^4{t})](\cos^2{t}+k\sin^2{t})^{-1.5}\end{eqnarray}
 and 
 \begin{eqnarray}
 \rho&=&\frac{3(k-1)^2\sin^2{t}\cos^2{t}}{\cos^2{t}+k\sin^2{t}}, \\ p&=&-\frac{3(k-1)^2\sin^2{t}\cos^2{t}}{\cos^2{t}+k\sin^2{t}}-\frac{2[k-(\cos^4{t}+2k\sin^2{t}\cos^2{t}+k^2\sin^4{t})]}{(\cos^2{t}+k\sin^2{t})^{1.5}}.\end{eqnarray}
The last equation is the parametric EoS corresponding to the 
Pinney equation (3.7) [if exactly to its solution (3.8)] which we can write in the usual form as
\begin{equation}
 p=-\rho-2(kN^{-3}-N).\end{equation} The corresponding EoS 
 parameter takes the form
 \begin{equation}
 \omega=-\frac{1}{3}-\frac{2k}{3(\cos^2{t}+k\sin^2{t})^{2}}.\end{equation}
  
  2) Let us we consider another derivation of the Pinney cosmology. 
  For the flat FRW metric (2.2), the  usual Einstein-Dirac equation has the form
	\begin{eqnarray}
	3H^2-\rho&=&0,\\ 
		2\dot{H}+3H^2+p&=&0,\\
				\dot{\psi}+1.5H\psi+i\gamma^0V_{\bar{\psi}}&=&0,\\ 
\dot{\bar{\psi}}+1.5H\bar{\psi}-iV_{\psi}\gamma^{0}&=&0,\\
	\dot{\rho}+3H(\rho+p)&=&0,
	\end{eqnarray} 
where    the kinetic term, the energy density  and  the pressure  take the form
  \begin{equation}
 Y=0.5i(\bar{\psi}\gamma^{0}\dot{\psi}-\dot{\bar{\psi}}\gamma^{0}\psi),\quad 
  \rho=V,\quad
p=uV_{u}-V,\quad u=\bar{\psi}\psi.
  \end{equation}
  Let the potential has the form
  \begin{equation}
 V=V_0+mu+3c^{\frac{2}{3}}\xi_1u^{-\frac{2}{3}}-3 c^{-\frac{2}{3}}\xi_2u^{\frac{2}{3}}-1.5c^{-\frac{4}{3}}\xi_3u^{\frac{4}{3}},
  \end{equation}
 where $\xi_j=\xi_j(t)$ are some function of $t$ and $V_0,m,c=consts$. In this case we have
 \begin{equation}
 \dot{Y}+3Hu(V_u+uV_{uu})=0,\quad \dot{V}+3HY=0,\quad u=ca^{-3}.
  \end{equation}
 As 
  \begin{equation}
 a_{\eta\eta}=-\frac{Y}{2a}\end{equation}
 for the potential (3.21), we get   that the scale factor satisfies the following equation
  \begin{equation}
a_{\eta\eta}=\xi_1(\eta)a+\xi_2(\eta)a^{-3}+\xi_4(\eta)a^{-4}+\xi_3(\eta)a^{-5},\end{equation}
 where $d\eta=adt, \quad \xi_4=0.5cm$. Consider particular cases. 
 
 i) Let $\xi_1=-\theta^2(\eta), \quad \xi_2=\kappa, \quad \xi_4=\xi_5=0$. Then the equation (3.24) becomes\begin{equation}
a_{\eta\eta}+\theta^2(\eta)a-\frac{\kappa}{a^{3}}=0.\end{equation}
It is the Pinney or Ermakov-Pinney equation. The general solution of Eq. (3.25) 
 is given by \cite{Pinney}
  \begin{equation}
a=\sqrt{Az_1^2+Bz_2^2+2Cz_1z_2},\end{equation}where $z_j(\eta)$ are linearly independent solutions to the equation
\begin{equation}
z_{j\eta\eta}+\theta^2(\eta)z_j=0.\end{equation}
Here the constants $A,B,C$ satisfy the constraint
\begin{equation}
AB-C^2=\kappa W^{-2}\end{equation}
 and 
 \begin{equation}
W\equiv z_1z_{2\eta}-z_2z_{1\eta}\end{equation}
is the Wronskian.

ii) We now consider the case: $\xi_1=-\theta^2_0=const, \quad \xi_2= \xi_4=0, \quad \xi_5=-\kappa$. Then Eq. (3.24) takes the form  \cite{Guha}
 \begin{equation}
a_{\eta\eta}+\theta^2_0a+\frac{\kappa}{a^{5}}=0.\end{equation}

\subsection{Schr$\ddot{o}$dinger cosmology}

We now reconstruct the FRW cosmological model induced by 
the linear Schr$\ddot{o}$dinger equation 
\begin{equation}
 \ddot{N}=uN+kN,\end{equation}
 where $u=u(t),\quad k=const$. Then the parametric EoS takes  the form
\begin{eqnarray}
		p&=&-3\dot{N}^2-2uN-2kN,\\
	\rho&=&3\dot{N}^2.
	\end{eqnarray}
 It is well-known that as $u=n(n-1)t^{-2}-k$, 
 Eq.(3.31) has  the following  particular solution 
 \begin{equation}
 N=\lambda t^n.\end{equation}
Then
 \begin{eqnarray}
 \rho&=&3\lambda^2 n^2t^{2(n-1)}, \\ p&=&-3\lambda^2 n^2t^{2(n-1)}-2n(n-1)\lambda t^{n-2}\end{eqnarray}
 or
\begin{equation}
 p=-\rho-2n(n-1)\lambda^{\frac{1}{n-1}}3^{\frac{2-n}{2(n-1)}}\rho^{\frac{n-2}{2(n-1)}}.
 \end{equation} The corresponding EoS 
 parameter takes the form
 \begin{equation}
 \omega=-1+\frac{2(n-1)}{3n\lambda t^{n}}.\end{equation}
 
 \subsection{Hypergeometric cosmology}

  i) We start with the equation
  \begin{equation}
 \ddot{N}=t^{-1}(1-t)^{-1}\{[(a+b+1)t-c]\dot{N}+abN\}.\end{equation}
 It is the hypergeometric differential equation. In this case, 
  the parametric EoS has the form
\begin{eqnarray}
		p&=&-3\dot{N}^2-2t^{-1}(1-t)^{-1}\{[(a+b+1)t-c]\dot{N}+abN\},\\
	\rho&=&3\dot{N}^2.
	\end{eqnarray}
The  solution of the equation (3.39) is the hypergeometric function 
 \begin{equation}N= \,_{2}F_{1}(a,b;c;t).
 \end{equation} Some particular solutions are:
  \begin{eqnarray}
 \ln(1+t)&=&t\,_{2}F_{1}(1,1;2;-t), \\
 (1-t)^{-a}&=&_{2}F_{1}(a,b;b;t),\\
 \arcsin{t}&=&t\,_{2}F_{1}(0.5,0.5;1.5;t^2),
 \end{eqnarray}
and so on. As an example, let us consider the solution (3.45) that is $N=\arcsin{t}$. 
 Then
\begin{equation}
 \dot{N}=(1-t^2)^{-0.5},\quad \ddot{N}=t(1-t^2)^{-1.5}.\end{equation}
 and 
 \begin{eqnarray}
 \rho&=&3(1-t^2)^{-1}, \\ p&=&-3(1-t^2)^{-1}-2t(1-t^2)^{-1.5}.\end{eqnarray}
 Hence we obtain
\begin{equation}
 p=-\rho-3^{-1.5}2t\rho^{1.5}\end{equation} 
 or
 \begin{equation}
 p=-\rho-3^{-1.5}2(\rho-3)^{0.5}\rho.\end{equation} For this case, 
 the  EoS 
 parameter takes the form
 \begin{equation}
 \omega=-1-\frac{2t}{\sqrt{1-t^2}}.\end{equation}
 
 ii) As the next example of the hypergeometric cosmology we can consider 
  models induced by elliptic integrals.  As an example, let us consider the complete elliptic integral of the first kind $K$:
  \begin{equation}
 K(t)=\int_0^1\frac{dz}{\sqrt{(1-z^2)(1-t^2z^2)}}.\end{equation}
 Assuming $N=K(t)$,  we can calculate all expressions for this case. But we 
 drop it as this case is the particular reduction of the model (3.42) due to of
 $N=K(t)=0.5\pi \,_{2}F_{1}(0.5,0.5;1;t)$.
 
 iii) The last example is the case when $N$ is equal to one of  incomplete elliptic integrals. For example, we can put   \begin{equation}
 N(t)=F(t; k)=\int_0^t\frac{dz}{\sqrt{(1-z^2)(1-k^2z^2)}}.\end{equation}
 Similarly to the previous cases we can also find all expressions to describe the cosmological model but omit it.
\subsection{Weierstrass cosmology}
\subsubsection{The Weierstrass gas}
One of the most interesting examples of Weierstrass cosmologies is the Weierstrass gas which has the following EoS
 \begin{equation}
 p=-B[\wp(\rho)]^{0.5}.\end{equation}
 The well-known Chaplygin gas model \cite{Kamenshchik}
  \begin{equation}
 p=-\frac{B}{\rho}\end{equation}
 is the limit (or particular case) of the Weierstrass gas when the Weierstrass function takes the form
  \begin{equation}
 \wp(\rho)=\rho^{-2}.\end{equation}
 
 \subsubsection{The generalized Weierstrass gas}
 The generalized Weierstrass gas corresponds to the  EoS
  \begin{equation}
 p=-B[\wp(\rho)]^{0.5\alpha}.\end{equation}
 Its limit (or the particular case) is the generalized Chaplygin gas \cite{Kamenshchik}
  \begin{equation}
 p=-\frac{B}{\rho^{\alpha}}\end{equation}
 that follows from  (3.56) and (3.57). 
 \subsubsection{The modified Weierstrass gas}
 At last we present the modified Weierstrass gas (MWG) model. Its EoS is given by
   \begin{equation}
 p=A\rho-B[\wp(\rho)]^{0.5\alpha}.\end{equation}
 The MWG is some generalization of the modified Chaplygin gas \cite{Benaoum}
   \begin{equation}
 p=A\rho-\frac{B}{\rho^{\alpha}}.\end{equation}
 Finally we would like to give the more general form of the MWG. Its EoS reads as
  \begin{equation}
 p=A[\wp(\rho)]^{-0.5}-B[\wp(\rho)]^{0.5\alpha}.\end{equation}
 
\section{Integrable FRW cosmological models}
After the previous section, we think that logically we are here in the position to make the next step, namely, to consider some integrable FRW cosmologies  (see e.g. \cite{Kuralay}-\cite{Nugmanova2}). Here we restrict ourselves
 to Painlev$\acute{e}$ cosmology that is cosmologies induced by the well-known Painlev$\acute{e}$ equations. These  six Painlev$\acute{e}$ equations (P$_I$ -- P$_{VI}$ equations) were first discovered about a hundred years ago by Painlev$\acute{e}$ and his colleagues in an investigation of nonlinear second-order ordinary differential equations. Recently, there has been considerable interest in the Painleve equations primarily due to the fact that they arise as reductions of the soliton equations which are solvable by Inverse Scattering Method (see e.g. \cite{Ablowitz}). Consequently, the Painlev$\acute{e}$ equations can be regarded as \textit{completely integrable equations} and possess solutions which can be expressed in terms of solutions of linear integral equations, despite being nonlinear equations.  The Painlev$\acute{e}$ equations may be thought of a nonlinear analogues of the classical special functions. They possess hierarchies of rational solutions and one-parameter families of solutions expressible in terms of the classical special functions, for special values of the parameters. Further the Painlev$\acute{e}$ equations admit symmetries under affine Weyl groups which are related to the associated Backlund transformations. Although first discovered from strictly mathematical considerations, the Painlev$\acute{e}$ equations have arisen in a variety of important physical applications including statistical mechanics, plasma physics, nonlinear waves, quantum gravity, quantum field theory, general relativity, nonlinear optics and fibre optics. In this section, we show that these  six Painlev$\acute{e}$ equations also can describe,  in particular, the accelerated expansion of the universe that is the dark energy.

 \subsection{Painlev$\acute{e}$ cosmology}
We work with  the N-form of the Einstein equations that is with the system (2.14)-(2.16).
 We now assume that $N$ satisfies one of Painlev$\acute{e}$
	equations. Consider examples [below $\alpha, \beta, \gamma, \delta, \kappa$ and $\mu$ are arbitrary constants]. 
	\subsubsection{P$_{I}$ - model}
	Let the parametric EoS has the form
\begin{eqnarray}
		p&=&-3\dot{N}^2-12N^2-2t,\\
	\rho&=&3\dot{N}^2.
	\end{eqnarray}
Then we can show that  $N$ satisfies the P$_{I}$ - equation \cite{Ablowitz}
\begin{equation}
 \ddot{N}=6N^2+t.\end{equation}
 It is well-known  that the P$_I$ - equation can be expressed as the compatibility condition 
  \begin{equation}
 A_{t}-B_{\lambda}+[A,B]=0\end{equation}
 of the following  linear system
 \begin{eqnarray}
		\frac{\partial \Phi}{\partial \lambda}&=&A\Phi,\\
	\frac{\partial \Phi}{\partial t}&=&B\Phi,
	\end{eqnarray}
 where $A=A(t,\lambda)$ and $B=B(t,\lambda)$  are matrices, $\lambda$ is a spectral parameter which 
  is independent of $t$. For the P$_I$ - equation, $A$ and $B$ have the form
  \begin{eqnarray}
A&=&(4\lambda^4+2N^2+t)\sigma_3-i(4\lambda^2N+2N^2+t)\sigma_2-(2\lambda\dot{N}+0.5\lambda^{-1})\sigma_1,\\
	B&=&(\lambda+\lambda^{-1}N)\sigma_3-i\lambda^{-1} N\sigma_2,
	\end{eqnarray}
where Pauli matrices are given by
$$\sigma_1 = \begin{pmatrix}
0&1\\
1&0
\end{pmatrix}, \quad 
\sigma_2 = \begin{pmatrix}
0&-i\\
i&0
\end{pmatrix}, \quad \sigma_3 = \begin{pmatrix}
1&0\\
0&-1\end{pmatrix}.
$$

\subsubsection{P$_{II}$ - model}
Our next example is the case when  $N$ satisfies the P$_{II}$ - equation
\begin{equation}
 \ddot{N}=2N^3+(t-t_0)N+\alpha,\end{equation}
 where $N=N(t,\alpha), \quad \alpha=const$. Then  the parametric EoS has the form
\begin{eqnarray}
		p&=&-3\dot{N}^2-4N^3-2(t-t_0)N-2\alpha,\\
	\rho&=&3\dot{N}^2.
	\end{eqnarray}
We note that the P$_{II}$ - equation (4.9) can written as the compatibility condition (4.4) of the linear system (4.5) -- (4.6) with
\begin{eqnarray}
A&=&-i(4\lambda^2+2N^2+t)\sigma_3-2\dot{N}\sigma_2+(4\lambda N-\alpha\lambda^{-1})\sigma_1,\\
	B&=&-i\lambda\sigma_3+ N\sigma_1.
	\end{eqnarray}
It is well-known that the P$_{II}$ - equation has the following particular solution \cite{Ablowitz}
 \begin{equation}
 N=N(t,1)=-(t-t_0)^{-1}.\end{equation}
Then
\begin{equation}
 \dot{N}=(t-t_0)^{-2}, \quad \ddot{N}=-2(t-t_0)^{-3}\end{equation}
 and 
 \begin{equation}
 \rho=3(t-t_0)^{-4}, \quad p=4(t-t_0)^{-3}-3(t-t_0)^{-4}.\end{equation}
 For this example, the  EoS and its parameter take the form
 \begin{equation}
 p=-\rho+4\left(\frac{\rho}{3}\right)^{0.75}\end{equation}
 and
 \begin{equation}
 \omega=-1+\frac{4}{3}(t-t_0).\end{equation}So if $t<t_0+0.5 \quad (t>t_0+0.5)$ then this P$_{II}$-model describes the accelerated (decelerated) expansion of the universe and the case  $t=t_0$  corresponds to the cosmological constant. 
 \subsubsection{P$_{III}$ - model}
Let the parametric EoS has the form
\begin{eqnarray}
		p&=&-3\dot{N}^2-2\left[\frac{1}{N}\dot{N}^2-\frac{1}{t}(\dot{N}-\alpha N^2-\beta)+\gamma N^3+\frac{\delta}{N}\right],\\
	\rho&=&3\dot{N}^2.
	\end{eqnarray}
Then $N$ satisfies the P$_{III}$ - equation \cite{Ablowitz}
\begin{equation}
 \ddot{N}=\frac{1}{N}\dot{N}^2-\frac{1}{t}(\dot{N}-\alpha N^2-\beta)+\gamma N^3+\frac{\delta}{N},\end{equation}
 where $N=N(t,\alpha,\beta,\gamma, \delta)$. 
	One of particular solution
 of 
 the  P$_{III}$ - equation  has the form \cite{Ablowitz}
 \begin{equation}
 N=N(t,\alpha,0,0, -\alpha \kappa^3)=\kappa t^{1/3}.\end{equation}
Then
\begin{equation}
 \dot{N}=\frac{\kappa}{3}t^{-2/3}, \quad \ddot{N}=
 -\frac{2\kappa}{9}t^{-5/3}\end{equation}
 and 
 \begin{equation}
 \rho=\frac{\kappa^2}{3} t^{-4/3}, \quad p=-\frac{\kappa^2}{3} t^{-4/3}+
 \frac{4\kappa}{9} t^{-5/3}.\end{equation}
 The corresponding EoS and its parameter take the form
 \begin{equation}
 p=-\rho+\frac{4}{\sqrt[4]{27\kappa^{6}}}\rho^{1.25}\end{equation}
 and
 \begin{equation}
 \omega=-1+\frac{4}{3\kappa \sqrt[3]{t}}.\end{equation}
 
 \subsubsection{P$_{IV}$ - model}
Let us consider the parametric EoS of  the form
\begin{eqnarray}
		p&=&-3\dot{N}^2-2\left[\frac{1}{N}\dot{N}^2-\frac{1}{t}(\dot{N}-\alpha N^2-\beta)+\gamma N^3+\frac{\delta}{N}\right],\\
	\rho&=&3\dot{N}^2.
	\end{eqnarray}
Then  for  $N$ we come to  the  equation
\begin{equation}
 \ddot{N}=\frac{1}{2N}\dot{N}^2+1.5N^3+4tN^2+2(t^2-\alpha)N+
 \frac{\delta}{N},\end{equation}
 which nothing but the P$_{IV}$ - equation. Here $N=N(t,\alpha, \delta)$. One of particular solution
 of 
 this equation has the form \cite{Ablowitz}
 \begin{equation}
 N=N(t,0, -2)=-2 t.\end{equation}
Then
\begin{equation}
 \dot{N}=-2, \quad \ddot{N}=0\end{equation}
 and 
 \begin{equation}
 \rho=12, \quad p=-12.\end{equation}
 The corresponding EoS and its parameter take the form
 \begin{equation}
 p=-\rho,\quad  \omega=-1.\end{equation}
  \subsubsection{P$_{V}$ - model}
Let $N$ is the solution of  the P$_{V}$ - equation \cite{Ablowitz}
\begin{equation}
 \ddot{N}=(\frac{1}{2N}+\frac{1}{N-1})\dot{N}^2
 -\frac{1}{t}(\dot{N}-\gamma  N)+t^{-2}(N-1)^2(\alpha N+\beta N^{-1})
 +\frac{\delta N(N+1)}{N-1},\end{equation}
 where $N=N(t,\alpha,\beta,\gamma, \delta)$. The corresponding parametric EoS is given
 by
\begin{eqnarray}
		p&=&-3\dot{N}^2-2\left[(\frac{1}{2N}+\frac{1}{N-1})\dot{N}^2
 -\frac{1}{t}(\dot{N}-\gamma  N)+t^{-2}(N-1)^2(\alpha N+\beta N^{-1})
 +I\right],  \\
	\rho&=&3\dot{N}^2,
	\end{eqnarray}
where $I=\delta N(N+1)(N-1)^{-1}$. It is well-known that Eq.(4.34) 
has the following  particular solution (see e.g. \cite{Ablowitz}
 and references therein)
 \begin{equation}
 N=N(t;0,0,\mu, -0.5\mu^2)=\kappa e^{\mu  t}.\end{equation}
Then
\begin{equation}
 \dot{N}=\kappa \mu e^{\mu  t}, \quad \ddot{N}=
 \kappa \mu^2 e^{\mu  t}\end{equation}
 and 
 \begin{equation}
 \rho=3\kappa^2 \mu^2 e^{2\mu  t}, \quad p=-3\kappa^2 \mu^2 e^{2\mu  t}
 -2\kappa \mu^2 e^{\mu  t}.\end{equation}
 The corresponding EoS and its parameter take the form
 \begin{equation}
 p=-\rho\mp 2\mu\sqrt{\frac{\rho}{3}}\end{equation}
 and
 \begin{equation}
 \omega=-1-\frac{2}{3\kappa}e^{-\mu t}.\end{equation}
 
  \subsubsection{P$_{VI}$ - model}
Our last example of integrable FRW cosmologies is the P$_{VI}$ - model. Its the parametric EoS is given by
\begin{eqnarray}
	p&=&-3\dot{N}^2-2\left[0.5\left(\frac{1}{N}+\frac{1}{N-1}+\frac{1}{N-t}\right)\dot{N}^2
 -\left(\frac{1}{t}+\frac{1}{t-1}+\frac{1}{N-t}\right)\dot{N}
 +J\right],\\
	\rho&=&3\dot{N}^2,
	\end{eqnarray}
where $J=t^{-2}(t-1)^{-2}N(N-1)(N-t)\left[\alpha+ \beta tN^{-2}+\gamma(t-1)(N-1)^{-2}+\delta t(t-1)(N-t)^{-2}\right]$. The corresponding nonlinear ODS is the P$_{VI}$ - equation
$$
 \ddot{N}=0.5\left(\frac{1}{N}+\frac{1}{N-1}+\frac{1}{N-t}\right)\dot{N}^2
 -\left(\frac{1}{t}+\frac{1}{t-1}+\frac{1}{N-t}\right)\dot{N}
 $$\begin{equation}+t^{-2}(t-1)^{-2}N(N-1)(N-t)\left[\alpha+ \beta tN^{-2}+\gamma(t-1)(N-1)^{-2}+\delta t(t-1)(N-t)^{-2}\right],\end{equation}
 where $N=N(t,\alpha,\beta,\gamma, \delta)$. Its  particular solution is
 \cite{Ablowitz}
 \begin{equation}
 N=N(t;0.5\kappa^2,-0.5\kappa^2,0.5\mu^2,0.5(1-\mu^2))=t^{0.5}.\end{equation}
Then
\begin{equation}
 \dot{N}=0.5t^{-0.5}, \quad \ddot{N}=
-0.25t^{-1.5}\end{equation}
 and 
 \begin{equation}
 \rho=\frac{3}{4}t^{-1}, \quad p=-\frac{3}{4}t^{-1}
 +0.5t^{-1.5}.\end{equation}
 The corresponding EoS and its parameter take the form
 \begin{equation}
 p=-\rho+\sqrt{\frac{16}{27}}\rho^{1.5}\end{equation}
 and
 \begin{equation}
 \omega=-1+\frac{2}{3}t^{-0.5}.\end{equation}
 \subsection{Hamiltonian structure}
 It is very important that all Painlev$\acute{e}$ equations  can be represented as Hamiltonian systems that is as (see e.g. \cite{Ablowitz} and references therein)
 \begin{eqnarray}
		\dot{q}&=&\frac{\partial F}{\partial r},\\
	\dot{r}&=&-\frac{\partial F}{\partial q},
	\end{eqnarray}
 where $F(q, r, t)$ is the (non-autonomous) Hamiltonian function. Consider some examples (see e.g. \cite{Ablowitz} and references therein).
 
 1) P$_{I}$-model.  In this case, $q,r,F$ read as 
 \begin{eqnarray}
		\dot{q}&=&r,\\
	\dot{r}&=&6q^2+t,\\
	F&=&0.5r^2-2q^3-tq.
	\end{eqnarray}
	
	 2) P$_{II}$-model.  In this case we have
 \begin{eqnarray}
		\dot{q}&=&r-q^2-0.5t,\\
	\dot{r}&=&2qr+\alpha+0.5,\\
	F&=&0.5r^2-(q^2+0.5t)r-(\alpha+0.5)q.
	\end{eqnarray}
	
	 3) P$_{III}$-model.  In this case we have
 \begin{eqnarray}
		t\dot{q}&=&2q^2r-k_2tq^2-(2\theta_1+1)q+k_1t,\\
	t\dot{r}&=&-2qr^2+2k_2tqr+(2\theta_1+1)r-k_2((\theta_1+\theta_2)t,\\
tF&=&q^2r^2-[k_2tq^2+(2\theta_1+1)q-k_1t]r+k_2(\theta_1+\theta_2)tq.
	\end{eqnarray}
 \section{$F(R)$ - gravity models induced by second-order ODEs}
 In this section, we consider $F(R)$ - gravity models induced by second-order ODSs. Some of these $F(R)$ - gravity models are integrable and others are nonintegrable. 
 \subsection{Integrable $F(R)$ - gravity models}
 Our aim in this subsection is to present a class of integrable $F(R)$ - gravity models for the FRW metric case. To do it, we use again Painlev$\acute{e}$ equations. Consider the action of $F(R)$ - gravity \cite{Odintsov1} (see also e.g. \cite{Odintsov2}-\cite{MR2}) 
 \begin{equation}
S=\int\sqrt{-g}d^4x[R+f(R)+L_m]. \end{equation}
 In the case of the FRW metric, the equations for the action (5.1) are given by
  \begin{equation}
 3H^2=\rho_{eff}, \quad 2\dot{H}+3H^2=-p_{eff}.\end{equation}
 Here 
  \begin{eqnarray}
 \rho_{eff}&=&-0.5f+3(H^2+\dot{H})f^{'}-18H(\ddot{H}+4H\dot{H})f^{''}, \\ p_{eff}&=&0.5f-(3H^2+\dot{H})f^{'}+6(\dddot{H}+6H\ddot{H}+4\dot{H}^2+8H^2\dot{H})f^{''}+
 +36(\ddot{H}+4H\dot{H})^2f^{'''},\end{eqnarray}
 where $f^{'}=df/dR$ etc. To construct integrable $f(R)$ - gravity models, 
 we  assume  that the function $f(R)$ satisfies some integrable ODE. As an example, we here demand that $f(R)$ is a solution of one of Painlev$\acute{e}$ equations.  Let us present these  equations.
 \\
 1) $F_{I}(R)$ - models. a) $F_{IA}(R)$ - model:
  \begin{equation}
 f^{''}=6f^2+R\end{equation}
 or
   \begin{equation}
 F^{''}=6F^2+R.\end{equation}
 b) $F_{IB}(R)$ - model. Note that instead of these two models we can consider the following ones
  $$
 \ddot{f}=6f^2+t$$
 or
  $$
 \ddot{F}=6F^2+t.$$ \\
 2) $F_{II}(R)$ - models. a) $F_{IIA}(R)$ - model:
 \begin{equation}
 f^{''}=2f^3+Rf+\alpha\end{equation}
 or
  \begin{equation}
 F^{''}=2F^3+RF+\alpha.\end{equation}
 b) $F_{IIB}(R)$ - model. The alternative models are given by
 $$
 \ddot{f}=2f^3+tf+\alpha
 $$
 or
  $$
 \ddot{F}=2F^3+tF+\alpha.
 $$
 \\
 3) $F_{III}(R)$ - models. a) $F_{IIIA}(R)$ - model:
 \begin{equation}
 f^{''}=\frac{1}{f}f^{'2}-\frac{1}{R}(f^{'}-\alpha f^2-\beta)+\gamma f^3+\frac{\delta}{f}\end{equation}
 or
  \begin{equation}
 F^{''}=\frac{1}{F}F^{'2}-\frac{1}{R}(F^{'}-\alpha F^2-\beta)+\gamma F^3+\frac{\delta}{F}.\end{equation}
  b) $F_{IIIB}(R)$ - model. The alternative models are given by
  $$
 \ddot{f}=\frac{1}{f}\dot{f}^{2}-\frac{1}{t}(\dot{f}-\alpha f^2-\beta)+\gamma f^3+\frac{\delta}{f}
 $$
 or
 $$
 \ddot{F}=\frac{1}{F}\dot{F}^{2}-\frac{1}{t}(\dot{F}-\alpha F^2-\beta)+\gamma F^3+\frac{\delta}{F}.
 $$
 \\
 4) $F_{IV}(R)$ - models. a) $F_{IVA}(R)$ - model:
 \begin{equation}
 f^{''}=\frac{1}{2f}f^{'2}+1.5f^3+4Rf^2+2(R^2-\alpha)f+
 \frac{\delta}{f}\end{equation}
 or
  \begin{equation}
 F^{''}=\frac{1}{2F}F^{'2}+1.5F^3+4RF^2+2(R^2-\alpha)F+
 \frac{\delta}{F}.\end{equation}
  b) $F_{IVB}(R)$ - model:
  $$
 \ddot{f}=\frac{1}{2f}\dot{f}^{2}+1.5f^3+4tf^2+2(t^2-\alpha)f+
 \frac{\delta}{f}
 $$
 or
 $$
 \ddot{F}=\frac{1}{2F}\dot{F}^{2}+1.5F^3+4tF^2+2(t^2-\alpha)F+
 \frac{\delta}{F}.
 $$
 \\
 5) $F_{V}(R)$ - models. a) $F_{VA}(R)$ - model:
 \begin{equation}
 f^{''}=(\frac{1}{2f}+\frac{1}{f-1})f^{'2}
 -\frac{1}{R}(f^{'}-\gamma  f)+R^{-2}(f-1)^2(\alpha f+\beta f^{-1})
 +\frac{\delta f(f+1)}{f-1}\end{equation}
 or
  \begin{equation}
 F^{''}=(\frac{1}{2F}+\frac{1}{F-1})F^{'2}
 -\frac{1}{R}(F^{'}-\gamma  F)+R^{-2}(F-1)^2(\alpha F+\beta F^{-1})
 +\frac{\delta F(F+1)}{F-1}.\end{equation}
 b) $F_{VB}(R)$ - model:
  $$
 \ddot{f}=(\frac{1}{2f}+\frac{1}{f-1})\dot{f}^{2}
 -\frac{1}{R}(\dot{f}-\gamma  f)+t^{-2}(f-1)^2(\alpha f+\beta f^{-1})
 +\frac{\delta f(f+1)}{f-1}
 $$
 or
 $$
 \ddot{F}=(\frac{1}{2F}+\frac{1}{F-1})\dot{F}^{2}
 -\frac{1}{R}(\dot{F}-\gamma  F)+t^{-2}(F-1)^2(\alpha F+\beta F^{-1})
 +\frac{\delta F(F+1)}{F-1}.
 $$\\
 6) $F_{VI}(R)$ - models. a) $F_{VIA}(R)$ - model:
 $$
 f^{''}=0.5\left(\frac{1}{f}+\frac{1}{f-1}+\frac{1}{f-R}\right)f^{'2}
 -\left(\frac{1}{R}+\frac{1}{R-1}+\frac{1}{f-R}\right)f^{'}
 $$\begin{equation}+R^{-2}(R-1)^{-2}f(f-1)(f-R)\left[\alpha+ \beta Rf^{-2}+\gamma(R-1)(f-1)^{-2}+\delta R(R-1)(f-R)^{-2}\right].\end{equation}
 Instead of this equation we can consider the following one
   $$
 F^{''}=0.5\left(\frac{1}{F}+\frac{1}{F-1}+\frac{1}{F-R}\right)F^{'2}
 -\left(\frac{1}{R}+\frac{1}{R-1}+\frac{1}{F-R}\right)F^{'}
 $$\begin{equation}+R^{-2}(R-1)^{-2}F(F-1)(F-R)\left[\alpha+ \beta RF^{-2}+\gamma(R-1)(F-1)^{-2}+\delta R(R-1)(F-R)^{-2}\right].\end{equation}
 b) $F_{VIB}(R)$ - model: $$
\ddot{f}=0.5\left(\frac{1}{f}+\frac{1}{f-1}+\frac{1}{f-t}\right)\dot{f}^{2}
 -\left(\frac{1}{t}+\frac{1}{t-1}+\frac{1}{f-t}\right)\dot{f}
 $$
 $$+t^{-2}(t-1)^{-2}f(f-1)(f-t)\left[\alpha+ \beta tf^{-2}+\gamma(t-1)(f-1)^{-2}+\delta t(t-1)(f-t)^{-2}\right].
 $$
 and
 $$
 \ddot{F}=0.5\left(\frac{1}{F}+\frac{1}{F-1}+\frac{1}{F-t}\right)\dot{F}^{2}
 -\left(\frac{1}{t}+\frac{1}{t-1}+\frac{1}{F-t}\right)\dot{F}
 $$
 $$+t^{-2}(t-1)^{-2}F(F-1)(F-t)\left[\alpha+ \beta tF^{-2}+\gamma(t-1)(F-1)^{-2}+\delta t(t-1)(F-t)^{-2}\right].
 $$In the above, $\alpha, \beta, \gamma, \delta$ are some constants and the function $F(R)$ corresponds to the action
  \begin{equation}
S=\int\sqrt{-g}d^4x[F(R)+L_m]. \end{equation}
  Some comments in order. All above presented cosmological $F_{J}(R)- $models  $(J=I, II, III, IV, V, VI)$ are integrable due to of integrability  of Painlev$\acute{e}$ equations. In particular, this means that these models admit $n-soliton$ solutions.  As an example, let us present here some exact solutions for the $F_{II}$ - cosmology given by the equation (5.7). It is well-known that Eq. (5.7) has the following rational solution
    \begin{equation}
f(R)\equiv f(R;n)=\frac{d}{dR}\left(\ln\left({\frac{E_{n-1}(R)}{E_{n}(R)}}\right)\right), \end{equation}
where the $E_{n}(R)$ are monic polynomials (coefficient of highest power of $R$ is $1$) and satisfy the following equation
 \begin{equation}
E_{n+1}(R)E_{N-1}(R)=RE_{n}^2(R)+4E_{N}^{'2}(R)-4E_{N}(R)E_{n}^{''}(R). \end{equation}
This equation has the following first solutions \cite{Ablowitz}
\begin{eqnarray}
 E_0(R)&=&1, \\ 
 E_1(R)&=&R,\\
 E_2(R)&=&R^3+4,\\
 E_3(R)&=&R^6+20R^3-80,\\
 E_4(R)&=&R^{10}+60R^7+11200R,\\
 E_5(R)&=&R^{15}+140R^{12}+2800R^{9}+78400R^6-313600R^3-6272000,\\
 E_6(R)&=&R^{21}+280R^{18}+18480R^{15}+627200R^{12}-17248000R^{9}+L,
 \end{eqnarray}
 where $L=1448832000R^6+193177600000R^3-38635520000$ and so on.  The corresponding expressions for the function $f(R)$ have the form \cite{Ablowitz}
 \begin{eqnarray}
 f(R)\equiv f(R;1)&=&-\frac{1}{R}, \\ 
  f(R)\equiv f(R;2)&=&\frac{1}{R}-\frac{3R^2}{R^3+4},\\
  f(R)\equiv f(R;3)&=&\frac{3R^2}{R^3+4}-\frac{6R^2(R^3+10)}{R^6+20R^3-80},\\
f(R)\equiv f(R;4)&=&-\frac{1}{R}+\frac{6R^2(R^3+10)}{R^6+20R^3-80}-\frac{9R^5(R^3+40)}{R^9+60R^6+11200}
 \end{eqnarray}
and so on.  It is also interesting to note that the function $f(R)$ can be expressed by the so-called $\tau$-function as \cite{Ablowitz}
  \begin{equation}
f(R)\equiv f(R;n)=\frac{d}{dR}\left(\ln\left({\frac{\tau_{n-1}(R)}{\tau_{n}(R)}}\right)\right), \end{equation}
where \begin{equation}
  \tau_n(R)=\begin{vmatrix} p_1(R) & p_3(R) & \cdot & p_{2n-1}(R) \\ 
  p_1^{'}(R) & p_3^{'}(R) & \cdot & p_{2n-1}^{'}(R) \\ 
  \vdots & \vdots & \ddots & \vdots \\ 
  p_1^{(n-1)}(R) & p_3^{(n-1)}(R) & \cdot & p_{2n-1}^{(n-1)}(R)  \end{vmatrix}.\end{equation}
 Here $p_{j}(R)$ are the polynomials defined by $p_{j}(R)=0$ for $j<0$, and
  \begin{equation}
e^{\lambda R-\frac{4}{3}\lambda^3}=\sum_{j=0}^{\infty}p_{j}(R)\lambda^j. \end{equation}
 \subsection{Nonintegrable $f(R)$ - gravity models}
 Obviously that some $F(R)$ - gravity models can be constructed by nonintegrable second-order ODEs. Let us consider examples. 
 
 i) Our  first example is  the hypergeometric differential equation (see e.g. \cite{Odintsov2})
 \begin{equation}
R(1-R)f^{''}+[c-(a+b+1)R]f^{'}-abf=0. \end{equation}
It has the solution $f(R)=\,_2F(a,b; c; R)$ which is the hypergeometric function. We can also consider the $t$-version of the equation
 \begin{equation}
t(1-t)\ddot{f}+[c-(a+b+1)t]\dot{f}-abf=0 \end{equation}
with the solution $f(t)=\,_2F(a,b; c; t)$.

ii)  Another example is the case when $f(R)$ satisfies the Pinney equation
 \begin{equation}
f^{''}+\xi_{1}(R)f+\frac{\xi_{2}(R)}{f^3}=0. \end{equation}
 If $\xi_1=1,\quad \xi_2=\kappa=const$, this equation has the following solution \cite{Haas}
 \begin{equation}
f(R)=\cos^2R+\kappa^2\sin^2R. \end{equation}
 Its  "$t$-form" is $f(t)=\cos^2t+\kappa^2\sin^2t$ which is  the solution of the  Pinney equation  \begin{equation}
\ddot{f}+\xi_{1}(t)f+\frac{\xi_{2}(t)}{f^3}=0 \end{equation}
 as $\xi_1=1,\quad \xi_2=\kappa=const$.
 
 iii) Let us we present one more example. Let the function $f$ satisfies the equation
  \begin{equation}
f^{''}=6f^2-0.5g_2, \quad (g_2=const)\end{equation}
  or
    \begin{equation}
\ddot{f}=6f^2-0.5g_2.\end{equation}These equations admit the following solutions
   \begin{equation}
f(R)=\wp(R) \end{equation}
and
   \begin{equation}
f(R)\equiv f(t)=\wp(t), \end{equation}
where  $\wp(R)$ and $\wp(t)$ are  the Weierstrass elliptic functions. 
 
 \section{$F(G)$ - gravity models induced by second-order ODEs}
 The next important modified gravity theory is  $F(G)$ gravity. Let us now  we extend results of the previous section to the $F(G)$ gravity case that is consider $F(G)$ - gravity models induced by second-order ODSs. As in the previous $F(R)$ gravity case, some of these $F(G)$ - gravity models are integrable and others are nonintegrable. Consider examples. 
 \subsection{Integrable $F(G)$ - gravity models}
 In this subsection, we present  a class of integrable $F(G)$ - gravity models for the FRW metric case using  again Painlev$\acute{e}$ equations. The action of $F(G)$ - gravity  we write as \cite{Nojiri4}-\cite{Cognola1} (see also e.g. \cite{MR1}-\cite{MR2}) 
 \begin{equation}
S=\int\sqrt{-g}d^4x[R+f(G)+L_m] \end{equation}
or
 \begin{equation}
S=\int\sqrt{-g}d^4x[F(G)+L_m]. \end{equation}
Here 
 \begin{equation}
G=R^2-4R_{\mu\nu}R^{\mu\nu}+R_{\mu\nu\sigma\tau}R^{\mu\nu\sigma\tau} \end{equation}
is the Gauss-Bonnet invariant which for the FRW metric takes the form
 \begin{equation}
G=24H^2(\dot{H}+H^2). \end{equation}
The FRW-equations for the action (6.1) are given by
  \begin{equation}
 3H^2=\rho_{G}+\rho_m, \quad 2\dot{H}+3H^2=-(p_{G}+p_m).\end{equation}
 Here 
  \begin{eqnarray}
 \rho_{G}&=&Gf_G-f-24H^3\dot{G})f_{GG}, \\ p_{G}&=&-\rho_G+8H^2\dot{G}^2f_{GGG}+E.\end{eqnarray}
 where $E=-192f_{GG}(4H^6\dot{H}-8H^3\dot{H}\ddot{H}-6H^2\dot{H}^3-H^4\dddot{H}-3H^5\ddot{H}-18H^4\dot{H}^2)$ and  $f_{G}=df/dG$ etc. To construct integrable $f(G)$ - gravity models, 
 we  assume  that the function $f(G)$ satisfies some integrable ODE, namely,  one of Painlev$\acute{e}$ equations.  We here just list such integrable models, the investigation of which leave for the future studies. Let us present these  equations.
 \\
 1) \textit{$F_{I}(G)$ - models}. \\
 a) $F_{IA}(G)$ - model:
  \begin{equation}
 f_{GG}=6f^2+G\end{equation}
 or
   \begin{equation}
 F_{GG}=6F^2+G.\end{equation}
 b) $F_{IB}(R)$ - model:
  $$
 \ddot{f}=6f^2+t$$
 or
  $$
 \ddot{F}=6F^2+t.$$ \\
 2) \textit{$F_{II}(G)$ - models}. \\
 a) $F_{IIA}(G)$ - model:
 \begin{equation}
 f_{GG}=2f^3+Gf+\alpha\end{equation}
 or
  \begin{equation}
 F_{GG}=2F^3+GF+\alpha.\end{equation}
 b) $F_{IIB}(G)$ - model:
 $$
 \ddot{f}=2f^3+tf+\alpha
 $$
 or
  $$
 \ddot{F}=2F^3+tF+\alpha.
 $$
 \\
 3) \textit{$F_{III}(G)$ - models}. \\
 a) $F_{IIIA}(G)$ - model:
 \begin{equation}
 f_{GG}=\frac{1}{f}f_{G}^{2}-\frac{1}{G}(f_{G}-\alpha f^2-\beta)+\gamma f^3+\frac{\delta}{f}\end{equation}
 or
  \begin{equation}
 F_{GG}=\frac{1}{F}F_{G}^{2}-\frac{1}{G}(F_{G}-\alpha F^2-\beta)+\gamma F^3+\frac{\delta}{F}.\end{equation}
  b) $F_{IIIB}(G)$ - model:
  $$
 \ddot{f}=\frac{1}{f}\dot{f}^{2}-\frac{1}{t}(\dot{f}-\alpha f^2-\beta)+\gamma f^3+\frac{\delta}{f}
 $$
 or
 $$
 \ddot{F}=\frac{1}{F}\dot{F}^{2}-\frac{1}{t}(\dot{F}-\alpha F^2-\beta)+\gamma F^3+\frac{\delta}{F}.
 $$
 \\
 4) \textit{$F_{IV}(G)$ - models}. \\
  a) $F_{IVA}(G)$ - model:
 \begin{equation}
 f_{GG}=\frac{1}{2f}f_{G}^{2}+1.5f^3+4Gf^2+2(G^2-\alpha)f+
 \frac{\delta}{f}\end{equation}
 or
  \begin{equation}
 F_{GG}=\frac{1}{2F}F_{G}^{2}+1.5F^3+4GF^2+2(G^2-\alpha)F+
 \frac{\delta}{F}.\end{equation}
  b) $F_{IVB}(G)$ - model:
  $$
 \ddot{f}=\frac{1}{2f}\dot{f}^{2}+1.5f^3+4tf^2+2(t^2-\alpha)f+
 \frac{\delta}{f}
 $$
 or
 $$
 \ddot{F}=\frac{1}{2F}\dot{F}^{2}+1.5F^3+4tF^2+2(t^2-\alpha)F+
 \frac{\delta}{F}.
 $$
 \\
 5) \textit{$F_{V}(G)$ - models}. \\ a) $F_{VA}(G)$ - model:
 \begin{equation}
 f_{GG}=(\frac{1}{2f}+\frac{1}{f-1})f_{G}^{2}
 -\frac{1}{G}(f_{G}-\gamma  f)+G^{-2}(f-1)^2(\alpha f+\beta f^{-1})
 +\frac{\delta f(f+1)}{f-1}\end{equation}
 or
  \begin{equation}
 F_{GG}=(\frac{1}{2F}+\frac{1}{F-1})F_{G}^{2}
 -\frac{1}{G}(F_{G}-\gamma  F)+G^{-2}(F-1)^2(\alpha F+\beta F^{-1})
 +\frac{\delta F(F+1)}{F-1}.\end{equation}
 b) $F_{VB}(G)$ - model:
  $$
 \ddot{f}=(\frac{1}{2f}+\frac{1}{f-1})\dot{f}^{2}
 -\frac{1}{R}(\dot{f}-\gamma  f)+t^{-2}(f-1)^2(\alpha f+\beta f^{-1})
 +\frac{\delta f(f+1)}{f-1}
 $$
 or
 $$
 \ddot{F}=(\frac{1}{2F}+\frac{1}{F-1})\dot{F}^{2}
 -\frac{1}{R}(\dot{F}-\gamma  F)+t^{-2}(F-1)^2(\alpha F+\beta F^{-1})
 +\frac{\delta F(F+1)}{F-1}.
 $$\\
 6) \textit{$F_{VI}(G)$ - models}. \\ a) $F_{VIA}(G)$ - model:
 $$
 f_{GG}=0.5\left(\frac{1}{f}+\frac{1}{f-1}+\frac{1}{f-G}\right)f_{G}^{2}
 -\left(\frac{1}{G}+\frac{1}{G-1}+\frac{1}{f-G}\right)f_{G}
 $$\begin{equation}+G^{-2}(G-1)^{-2}f(f-1)(f-G)\left[\alpha+ \beta Gf^{-2}+\gamma(G-1)(f-1)^{-2}+\delta G(G-1)(f-G)^{-2}\right]\end{equation}
 or
   $$
 F_{GG}=0.5\left(\frac{1}{F}+\frac{1}{F-1}+\frac{1}{F-G}\right)F_{G}^{2}
 -\left(\frac{1}{G}+\frac{1}{G-1}+\frac{1}{F-G}\right)F_{G}
 $$\begin{equation}+G^{-2}(G-1)^{-2}F(F-1)(F-G)\left[\alpha+ \beta GF^{-2}+\gamma(G-1)(F-1)^{-2}+\delta G(G-1)(F-G)^{-2}\right].\end{equation}
 b) $F_{VIB}(G)$ - model: $$
\ddot{f}=0.5\left(\frac{1}{f}+\frac{1}{f-1}+\frac{1}{f-t}\right)\dot{f}^{2}
 -\left(\frac{1}{t}+\frac{1}{t-1}+\frac{1}{f-t}\right)\dot{f}
 $$
 $$+t^{-2}(t-1)^{-2}f(f-1)(f-t)\left[\alpha+ \beta tf^{-2}+\gamma(t-1)(f-1)^{-2}+\delta t(t-1)(f-t)^{-2}\right].
 $$
 and
 $$
 \ddot{F}=0.5\left(\frac{1}{F}+\frac{1}{F-1}+\frac{1}{F-t}\right)\dot{F}^{2}
 -\left(\frac{1}{t}+\frac{1}{t-1}+\frac{1}{F-t}\right)\dot{F}
 $$
 $$+t^{-2}(t-1)^{-2}F(F-1)(F-t)\left[\alpha+ \beta tF^{-2}+\gamma(t-1)(F-1)^{-2}+\delta t(t-1)(F-t)^{-2}\right].
 $$
 So all these models are integrable that is they admit n-soliton solutions, infinite number commuting integrals, Lax representations and so on. 
 \subsection{Nonintegrable $f(G)$ - gravity models}
 We now briefly list some nonintegrable $F(G)$ - gravity models which  can be constructed by nonintegrable second-order ODEs. Let us consider examples. 
 \\
 1) The hypergeometric differential equation (see e.g. \cite{Odintsov2})
 \begin{equation}
G(1-G)f_{GG}+[c-(a+b+1)G]f_{G}-abf=0 \end{equation}
or
 \begin{equation}
t(1-t)\ddot{f}+[c-(a+b+1)t]\dot{f}-abf=0. \end{equation}
Note that these equations have the following  solutions $f(G)=\,_2F(a,b; c; G)$ and $f(G)\equiv f(t)=\,_2F(a,b; c; t)$, respectively. 
\\
2)  The Pinney equation
 \begin{equation}
f_{GG}+\xi_{1}(G)f+\frac{\xi_{2}(G)}{f^3}=0 \end{equation}
 or
  \begin{equation}
\ddot{f}+\xi_{1}(t)f+\frac{\xi_{2}(t)}{f^3}=0. \end{equation}If $\xi_1=1,\quad \xi_2=\kappa=const$, these equations have the following solutions \cite{Haas}
 \begin{equation}
f(G)=\cos^2G+\kappa^2\sin^2G \end{equation}
 and  \begin{equation}
f(G)\equiv f(t)=\cos^2t+\kappa^2\sin^2t, \end{equation}
 respectively.
 \\
 3) Our last models are given by
  \begin{equation}
f_{GG}=6f^2-0.5g_2, \quad (g_2=const) \end{equation}
 or
  \begin{equation}
\ddot{f}=6f^2-0.5g_2 \end{equation}
 which admit the following solutions 
   \begin{equation}
f(G)=\wp(G) \end{equation}
  and
   \begin{equation}
f(G)\equiv f(t)=\wp(t) \end{equation}
respectively. Here $\wp(G), \wp(t)$ are  the Weierstrass elliptic functions. 
 
 \section{Scalar field description }
As is well-known scalar fields play an essential role in modern cosmology since they are possible candidates for the role of the inflaton field driving inflation in the early universe and of the dark energy substance responsible for the present cosmic acceleration. A way to describe the above presented cosmological models from field theoretical point of view is  to introduce a scalar field $\phi$ and self-interacting potential $U(\phi)$ with the following Lagrangian:
\begin{equation}
L_m=0.5\dot{\phi}^2-U(\phi). \end{equation}
The corresponding energy density and pressure are given by
\begin{equation}
\rho=0.5\dot{\phi}^2+U(\phi), \quad p=0.5\dot{\phi}^2-U(\phi).\end{equation}
Consider the $P_{II}$-cosmology (subsubsection 4.1.2). Then we get that
\begin{equation}
\phi=\int\sqrt{-2(2N^3+tN+\alpha)}dt,  \quad U=3\dot{N}^2+2N^3+tN+\alpha.\end{equation}
Hence  for the particular solution (4.7) we obtain\begin{equation}
\phi=\phi_0-4t^{-0.5},  \quad U=3t^{-4}-2t^{-3}.\end{equation}
It  gives
\begin{equation}
U=2^{-16}3(\phi-\phi_0)^{8}-2^{-11}(\phi-\phi_0)^{6}.\end{equation}
Similarly we can find the expressions of $\phi$ and its self-interaction potentials for the other examples considered above.
\section{Two-dimensional generalizations}
The above considered FRW models are one-dimensional. Here we present  their two-dimensional generalizations following of  \cite{Ablowitz}. 

a) As an example, let us consider 
 the P$_{II}$-model.
We introduce a new function $v(y,\mu)$ as
 \begin{equation}
v(y,\mu)=\frac{1}{\sqrt[3]{3\mu}}N(t),\end{equation}
where $N(t)$ is the solution of the P$_{II}$ - equation (4.9), $y=t\sqrt[3]{3\mu}$ is a new "time"-coordinate, $\mu$ is some physical parameter. Here as example, we take $\mu=\Lambda$, where $\Lambda$ is the cosmological constant. Then the function $v=v(y,\Lambda)$ obeys the following equation \cite{Ablowitz}
 \begin{equation}
v_{\Lambda}-6v^2v_y+v_{yyy}=0.\end{equation}
It is the famous modified Korteweg–de Vries (mKdV) equation which is integrable.

b) Similarly, we can construct two-dimensional generalizations of the other models. For example, let $v$ has the form
 \begin{equation}
v(y,\Lambda)=\frac{1}{\sqrt[3]{3\Lambda}}f(R),\end{equation}
where $f(R)$ is the solution of the $F_{II}$ - equation (5.7), $y=R\sqrt[3]{3\Lambda}$.
 Then the function $v=v(y,\Lambda)$ again satisfies  the equation (8.2). Next, if we take 
   $v$  in the form
 \begin{equation}
v(y,\Lambda)=-\frac{1}{\sqrt[3]{9\Lambda^2}}[N^{'}(t)+N^{2}(t)],\end{equation}
 where  $N(t)$ is again the solution of the P$_{II}$ - equation (4.9), then the function $v$ satisfies the  Korteweg–de Vries (KdV) equation \cite{Ablowitz}
  \begin{equation}
v_{\Lambda}+6vv_y+v_{yyy}=0,\end{equation}
which is also integrable. For the $F_{II}$ - model (5.7), $v$ is given by
 \begin{equation}
v(y,\Lambda)=-\frac{1}{\sqrt[3]{9\Lambda^2}}[f^{'}(R)+f^{2}(R)]\end{equation}
and $y=R\sqrt[3]{3\Lambda}$, where $f(R)$ is the solution of the equation (5.7).

c) Now we consider the two-dimensional generalization of the equations (4.21) and (5.9). Let $y=t\mu^{-1}$ and $N(t)$ is the solution of the P$_{III}$-equation (4.21) with $\alpha=-\beta=0.5, \quad \gamma=\delta=0$. Then the function 
 \begin{equation}
v(y,\mu)=i\ln{N(t)}\end{equation}
satisfies the sine-Gordon equation \cite{Ablowitz} (see also \cite{Dzhun2})
\begin{equation}
v_{y\mu}=\sin{v}.\end{equation}
If $y=R\mu^{-1}$ and $v(y,\mu)=i\ln{f(R)}$  then $f(R)$ satisfies  the equation (5.9).
\section{Conclusion}
In this work we have considered some  FRW cosmological models with the parametric EoS, where the e-folding $N$ plays the role of the parameter.  To construct the cosmological model in the explicit form we demand that $N$ is the solution of some linear or  nonlinear second-order ODSs. If such ODS are integrable then the corresponding FRW models are also integrable. It means that such models admit all ingredients of integrable systems such as n-soliton solutions, commuting integrals, Lax representations etc. 

Here we have constructed integrable FRW models induced by Painlev$\acute{e}$ equations. Then we have extended our results for modified $F(R)$ and $F(G)$ gravity theories. We have obtained the explicit forms of $f(R)$ and $f(G)$ functions for a flat FRW universe filled by some exotic fluid with the parametric EoS. We have then described one of considered models as FRW cosmological model having a scalar field and found its self-interacting potential. Note that all FRW models considered above are one-dimensional. Also  we have discussed the two-dimensional extensions of some models. In this case, as a second coordinate we take some physical parameter $\mu$, for example, the cosmological constant $\Lambda$. 

The main idea of our work is the reconstruction some gravitational models  using the  solutions of some differential equations (in our case, second-order ODSs). Here we would like to note that this approach works not only for second-order ODSs and can be extended to other ODSs and even to partial differential equations.  Let us briefly demonstrate this possibility. As an example, we can consider the case when $N$ is the solution of the following equation
 \begin{equation}
\dot{N}^2=4N^3-g_2N-g_3, \end{equation}
where $g_2, g_3$ are some constants. Its solution is $N(t)=\wp(t)$  that is   the Weierstrass elliptic function. The corresponding EoS is given by
 \begin{equation}
p=-\rho-12\wp^2(t)+g_2.\end{equation} The $f$-versions of Eq.(9.1) are $\dot{f}^2=4f^3-g_2f-g_3$, $f_R^2=4f^3-g_2f-g_3$ and $f_{G}^2=4f^3-g_2f-g_3$. Another interesting models follow from the  Bernoulli  equation:  \begin{equation}
\dot{N}=q_1(t)N^n+q_2(t)N. \end{equation}
 For example, if $n=2, q_1=-t^2, q_2=2t^{-1}$ then the Bernoulli equation has the solution $N=t^2(0.2t^5+const)^{-1}$.  The $f$-versions of this solution are  $f(R)=R^2(0.2R^5+const)^{-1}$ and  $f(G)=G^2(0.2G^5+const)^{-1}$. Finally we  would like to note that it would be interesting to make relation with 
viable models of $F(R)$ and $F(G)$ gravities unifying inflation with dark energy (see e.g. \cite{Nojiri5}-\cite{Elizalde2}). This important question will be the subject of the separate investigation. 

\section*{Acknowledgments}
We would like to thank the anonymous referee for providing us with constructive comments and suggestions to improve this work. One of the authors (R.M.) thanks   D. Singleton  and Department of Physics, California State University Fresno  for their hospitality during his one year visit (October, 2010 -- October, 2011).

 \end{document}